# A better approach to diagnose retinal diseases: Combining our Segmentation-based Vascular Enhancement with deep learning features

Yuzhuo Chen[1#*], Zetong Chen[1#], Yuanyuan Liu[1*]

[1] School of Control Science and Engineering, Shandong University, Jinan City, Shandong Province, 250061, China

#These authors contributed to the work equally and should be regarded as co-first authors.
***Corresponding authors**: Yuanyuan Liu

Yuanyuan Liu
School of Control Science and Engineering, Shandong University
Address: Qianfoshan Campus, Shandong University, 17923 Jingshi Road, Jinan City, Shandong Province, 250061, China
Telephone number: +86-13964199623
E-mail: liuyy@sdu.edu.cn

## Abbreviations

*SVE, Segmentation-based Vascular Enhancement; DL, Deep learning; ML, Machine learning; TL, Transfer Learning; ROC, Receiver operating characteristic; AUC, Area under curve; ROC, Receiver Operating Characteristic; AUC, the Area Under the Curve; TPR, True Positive Rate FPR, False Positive Rate;TP, True Positives; FN, False Negatives; FP, False Positives; TN, True Negatives*

# Abstract

Abnormalities in retinal fundus images may indicate certain pathologies such as diabetic retinopathy, hypertension, stroke, glaucoma, retinal macular edema, venous occlusion, and atherosclerosis, making the study and analysis of retinal images of great significance. In conventional medicine, the diagnosis of retina-related diseases relies on a physician's subjective assessment of the retinal fundus images, which is a time-consuming process and the accuracy is highly dependent on the physician's subjective experience. To this end, this paper proposes a fast, objective, and accurate method for the diagnosis of diseases related to retinal fundus images. This method is a multiclassification study of normal samples and 13 categories of disease samples on the STARE database, with a test set accuracy of 99.96%. Compared with other studies, our method achieved the highest accuracy. This study innovatively propose Segmentation-based Vascular Enhancement(SVE). After comparing the classification performances of the deep learning models of SVE images, original images and Smooth Grad-CAM ++ images, we extracted the deep learning features and traditional features of the SVE images and input them into nine meta learners for classification. The results shows that our proposed UNet-SVE-VGG-MLP model has the optimal performance for classifying diseases related to retinal fundus images on the STARE database, with a overall accuracy of 99.96% and a weighted AUC of 99.98% for the 14 categories on test dataset. This method can be used to realize rapid, objective, and accurate classification and diagnosis of retinal fundus image related diseases.

# 1. Introduction

Over the past five years, large population-based studies have shown that morphological and topographical changes in retinal blood vessels may predict certain pathologies such as diabetic retinopathy[1], hypertension[2], stroke[3], glaucoma[4], macular edema[5], venous occlusion, and arteriosclerosis[6]. In clinical diagnosis, the diagnosis of retina-related diseases relies on the physician's subjective assessment of the fundus retina. Not only does this process take a lot of time and effort, but the accuracy of the diagnosis depends heavily on the subjective experience of the doctor. With the development of artificial intelligence, a large number of computer-aided diagnosis (CAD) models based on machine learning have been developed. it has become a trend to rely on advanced computer technology to process and analyze retinal patient data, and to establish objective machine learning models and deep learning models to assist physicians in diagnosis[7][8][9].

In the study on retinal disease classification using machine learning models with deep learning, Rahim et al. [10]achieve an accuracy of 0.633 vs. 0.772 for the detection of diabetic retinopathy and macular degeneration, respectively, utilizing an "on-the-fly" (on-the-fly) data augmentation technique as well as a DCNN model. Penget al.[11] use DeepSeeNet to classify patients automatically by the AREDS Simplified Severity Scale (score 0-5) using bilateral CFP, which performed better on patient-based classification than retinal specialists. Chang al.[12] use Adversarial examples(AE) and gradient-weighted class activation mapping (GradCAM) to train DLMs with highest AUC 0.99 and AE provide better interpretability than GradCAM with board-trained glaucoma specialists rating Sabanayagam et al.[13] develop an artificially intelligent deep learning algorithm (DLA) for detecting chronic kidney disease (CKD) using retinal images data from three population-based, multi-ethnic, cross-sectional studies in Singapore and China, where hybrid DLA achieved an AUC of 0.938.Tong et al. [14]employ ResNet and Faster-RCNN networks to classify the severity of retinopathy of prematurity from fundus images, with a classification accuracy of 0.903.Das et al.[15] propose a Deep Multiscale Fusion Convolutional Neural Network (DMF-CNN) that achieved an overall accuracy of 0.960 and 0.996 on the UCSD and NEH datasets, respectively. Keenan et al.[16] use DeepLensNet to perform automated diagnosis and quantitative classification of age-related cataract from anterior segment photographs, the mean MSE for DeepLensNet was 0.23 for NS, 13.1 for CLO, and 16.6 for PSC. Almustafa et al.[17] utilize the data enhancement strategy of rotation and mirror flipping and used networks such as EfficientNet and inception V2 for fundus retinal images classification, achieving a maximum accuracy of 0.984 on the STARE database.

In addition to directly using deep learning models to classify retinal fundus images, using deep learning networks to extract features in the image and establishing a meta learner to classify the features is a common approach. Pratap et al. [18]use a CNN network to achieve automatic classification of the different stages of cataracts and use pre-trained CNNs for feature extraction then apply the extracted features to a Support Vector Machine (SVM) classifier, achieving an accuracy of 0.929. Gan et al.[19] introduce a multi-feature fusion to detect macular edema (ME), fusing traditional omics features with deep learning features from alexnet,inception_v3, and the fused features achieved an accuracy of 0.938 on SVM. Gayathri et al.[20] propose a novel CNN model for extracting features from retinal fundus images for the detection of diabetic retinopathy (DR), and the extracted features achieved an accuracy of 0.998 in J48 Decision Tree Classifier. Although there have been many studies using deep learning networks to classify retina-related diseases using fundus images, current studies often use retinal images directly for classification and do not utilize the importance of vascular features for retina-related diseases.

In order to enhance the model's attention to blood vessels, a preprocessing method of Segmentation-based Vascular Enhancement (SVE) was proposed. The effectiveness of SVE was verified by comparing with other preprocessing methods. In order to solve the problem of the imbalance in the number of different categories and the overfitting problem of the deep learning model, this study used a variety of data augmentation strategies. Firstly, the image is rotated at different angles, and then the Gaussian noise is superimposed on the image and then cutmix the image of the same category. Then the random crop is applied to each image. After experimentally verifying the best preprocessing method, this study extracted the deep learning features and traditional features based on the preprocessed images, and fed them into eight different one-dimensional models for classification performance comparison. Among them, the SVE-VGG-MLP model combined with SVE preprocessing, VGG feature extractor and MLP decoder had the best classification performance for retinal fundus image-related diseases. This method can be used to further develop diagnostic equipment for retinal fundus image-related diseases in clinical practice, so as to achieve rapid, objective and accurate classification and diagnosis of retinal fundus image-related diseases. Our main contributions are as follows:

1) We Proposed a novel preprocessing method named Segmentation-based Vascular Enhancement(SVE) which can enhances the attention of models to vascular features for retina-related diseases.

2) We Compared our preprocessing methods of SVE with Smooth Grad-CAM ++ and deep learning features with traditional features.

3) Our proposed model achieves the state-of-the-art in STARE dataset.

The structure of our SVE-VGG-MLP model is shown in Figure.1.

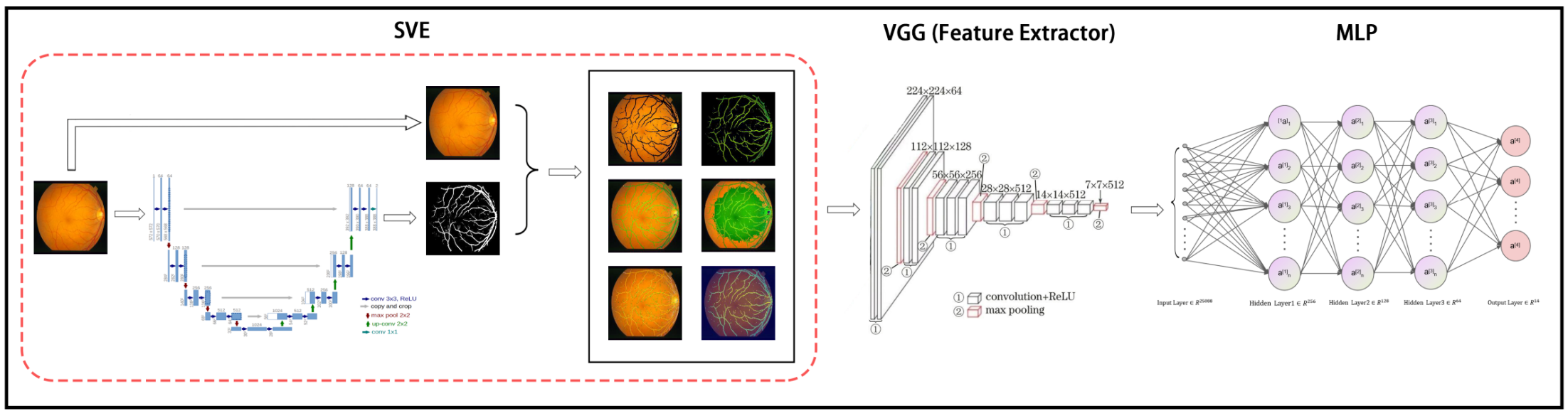

**Figure.1. The structure diagram of SVE-VGG-MLP.**

## 2. Methods

### *2.1. Sample preparation*

A total of 113 retinal fundas images corresponding to manually segmented blood vessel images were collected, including 28 from the CHASE DB1 database[21], 20 from the DRIVE database[22], 45 from the HRF database[23], and 20 from the STARE database[24] (Table.1).

A total of 397 retinal images corresponding to disease categories labeled by a medical professional on the STARE database were collected. However, since there are various diseases corresponding to label 14 in the STARE database, which cannot be visually discriminated, we deleted 92 cases of images corresponding to label 14 on the STARE database, leaving a total of 305 cases, including 39 cases of which were normal and 267 cases of 13 categories of samples suffering from diseases. The number of samples in each category with the given Label is shown in Table.2. In this study, the training set, validation set and test set were divided using the ratio of 6:2:2.

| Database | Number of samples |
|---|---|
| CHASE DB1 | 28 |
| DRIVE | 20 |
| HRF | 45 |
| STARE | 20 |

**Table.1. The samples for segmentation.**

| Disease | Abbreviation used | Number of samples | Label |
|---|---|---|---|
| None (Normal) | None (Normal) | 39 | 0 |
| Hollenhorst Emboli | Emboli | 6 | 1 |
| Branch Retinal Artery Occlusion | BRAO | 6 | 2 |
| Cilio-Retinal Artery Occlusion | CRAO | 9 | 3 |
| Branch Retinal Vein Occlusion | BRVO | 11 | 4 |
| Central Retinal Vein Occlusion | CRVO | 21 | 5 |
| Hemi-Central Retinal Vein Occlusion | Hemi-CRVO | 12 | 6 |
| Background Diabetic Retinopathy | BDR/NPDR | 57 | 7 |
| Proliferative Diabetic Retinopathy | PDR | 23 | 8 |
| Arteriosclerotic Retinopathy | ASR | 22 | 9 |
| Hypertensive Retinopathy | HTR | 25 | 10 |
| Coat's | None | 14 | 11 |
| Macroaneurism | None | 8 | 12 |
| Choroidal Neovascularization | CNV | 52 | 13 |

**Table.2. The samples for classification.**

## *2.2. Data augmentation*

This study employed diverse data augmentation strategies, including rotating different types of samples at different angles, mirror flipping, overlaying Gaussian noise with a mean of 0 and a standard deviation of 10 on the images, performing cutmix on images of the same category, and performing Random Crop on each image during training.

Rotation[25] is the process of rotating an image by a certain angle around the center point to generate new training samples. Performing smaller and more rotations on a smaller number of types of samples can alleviate the problem of class imbalance. Mirror flipping[26] is flipping an image horizontally or vertically to generate new training samples. Gaussian noise[27] is a type of random noise that follows a normal distribution, Which can be added to input data for data augmentation. Cutmix[28] is the process of cutting out a portion of an area and randomly filling it with pixel values from other data in the training set. Random crop[29] refers to cropping at random positions in an image for a specified size.

Rotation and mirror reversal can help models better learn features from different angles and directions, improving their recognition ability for objects in different directions. After adding Gaussian noise, the model is forced to learn features that are robust to small changes in the input, which improve the robustness and generalization ability of model. Cutmix can further enhance the localization ability of the model by requiring model to recognize the object from the local view and adding the information of other samples to the cut region. Random Crop can establish the weight relationship between each factor feature and the corresponding category, weaken the weight of background or noise factors, and make the model insensitive to missing values, so as to produce better learning effect and increase model stability.

Through diverse data augmentation strategies, this study alleviated the model preference problem caused by class imbalance, improved the robustness of the model, reduced the overfitting problem of the model, and improved the final performance of the model.

## *2.3. Image Preprocessing*

### *2.3.1. Segmentation-based Vascular Enhancement (SVE)*

When doctors diagnose different retina-related diseases through retinal fundus images, observing the structure and appearance of blood vessels in the image is often an significant step in diagnosis. To improve the model's ability to capture vascular features, this study proposed Segmentation-based Vascular Enhancement (SVE) to enhance the vascular area in the retinal fundus image based on the results of segmentation model.

In recent years, some retinal blood vessel segmentation networks with better segmentation results have been proposed in DRIVE, CHASE-DB1 and STARE databases. Deep Fully Convolutional Neural Networks （FCN）[30], IterNet[31], SA-UNet[32], Study Group Learning[33], RV-GAN[34], FR-UNet[35], IDmUNet[36] , Squeeze Excitation Residual UNet (SER-UNet)[37] and Hybrid Attention Fusion U-Net Model (HAU-Net)[38] are some of typical networks.

In this study, we trained RV-GAN, FR-UNet, R2U-Net and UNet models in the joint database of CHASE DB1, DRIVE, HRF and STARE, which have been performing better in the task of retinal blood vessel segmentation in recent years. After comparing segmentation results of the above four models, we found the UNet performance best on test data set. So, we chose UNet segmentation model to segment vessel of all original image on STARE database.

In this study, several SVE strategies were tried, and the SVE images of different strategies are shown in Figure.2. After comparing the classification results of different strategies of SVE, the Weighted blood vessel + Background strategy with the best classification results was chosen in this study, as follows:

The RGB channel of the original image is transferred to the HSI channel, keeping the values of the H and S channels unchanged and transforming only the values of the I channel.

$$I_background = I * (1 - mask) \tag{1}$$

$$I_enhanced = I + mask \times 255 \times weight \tag{2}$$

$$I_enhanced_normalized = \frac{I_enhanced - min(I_enhanced)}{max(I_enhanced) - min(I_enhanced)} \times 255 \tag{3}$$

$$I_enhanced_normalized_vessel = I_enhanced_normalized \times mask \tag{4}$$

$$I_new = I_background + I_enhanced_normalized_vessel \tag{5}$$

Where, I_background is the background of the I channel excluding blood vessels, I_enhanced is the matrix of the I channel after enhancement, I is the original image I channel value matrix. mask is the binary mask of the segmented image, there are only 0 and 1, so it will be multiplied by 255 to get the eight-bit binary image, I_enhanced_normalized is the I_enhanced image after normalization to 0-255.Weight is the weight of the enhanced image, the larger the weight is, the more obvious the enhancement effect is. After comparison, this study adopts weight=0.2. I_enhanced_normalized_vessel is the vessel extracted from I_enhanced_normalized. I_new is the final obtained enhanced image.

Then, in this study, we fused the H, S, and I_new channel data and transformed them back to the R, G, and B channels to obtain the blood vessel enhancement image.

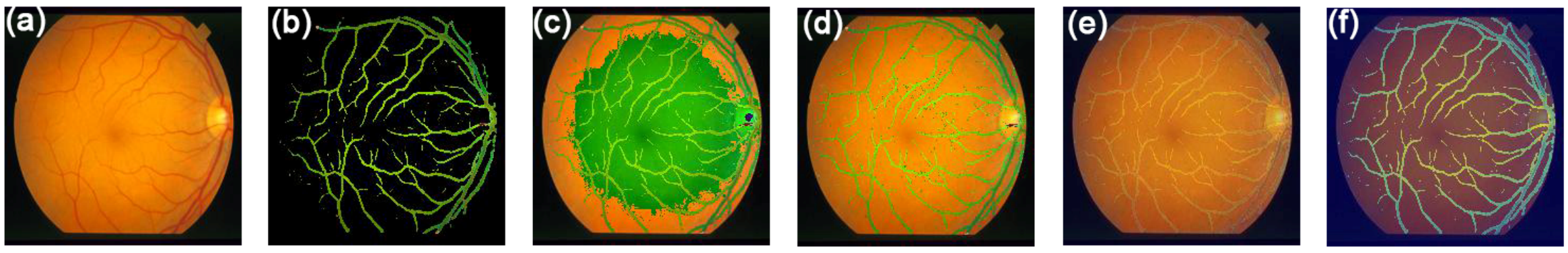

**Figure.2. The origin image and SVE images under different enhancement strategies.** (a)Origin. (b)Blood vessel. (c)Weighted blood vessel + Origin. (d)Weighted blood vessel + Background. (e)Gamma-transformed blood vessel + Origin. (f)Heatmap-transformed blood vessel + Origin.

### *2.3.2. Smooth Grad-CAM ++*

Smooth Gradient based Class Activation Maps plus plus (Smooth Grad-CAM++)[39] is a technology

combined with Smooth Grad and Grad-CAM++, which provides the capability of either visualizing a layer, subset of feature maps, or subset of neurons within a feature map at each instance at the inference level (model prediction process). Compared to other CAM methods, Smooth Grad-CAM++ introduces a gradient smoothing technique that produces visually sharper Maps and better localization of objects in a given input image.

In this study, Smooth Grad-CAM++ was used to extract the Activation Map from the original image and superimposed on the original image to obtain the Smooth Grad-CAM++ activation image. Then we used the Smooth Grad-CAM++ images as an input to train ViT, VGG, ResNet models and to evaluate the effect of this preprocessing using classification metrics

## *2.4. Feature Extraction*

### *2.4.1. Traditional features extraction*

Histogram of Oriented Gradients(HOG)[40] is a descriptor used to characterize the local gradient direction and gradient intensity distribution of an image, which was firstly used to solve the human target detection. HOG represents the structural features of the edges, so it can describe the local shape information, and the quantization of the position and direction can suppress the effects of translation and rotation to a certain extent.

Local Binary Pattern(LBP)[41] is an algorithm used to characterize the local texture of an image, reflecting the texture changes around the image pixels, with the advantages of rotational invariance, gray scale invariance (no effect of illumination changes), and low computational complexity. Its robustness to illumination changes is good because it focuses on local pixel relative relationships rather than absolute gray values.

In this study, gradient features in SVE images were extracted using HOG and LBP algorithms, and their classification results based meta learner are compared with deep learning feature classification results.

### *2.4.2. Deep learning features extraction*

Deep learning features [42] is the features of a sample through an intermediate level (a certain hidden layer) in a pre-trained neural network. Deep learning models can gradually construct feature representations through multiple levels of abstraction hierarchies to obtain higher-level semantic features from the underlying features. Since deep learning models can automatically learn from a large amount of data, they usually have better generalization ability when the training samples are sufficient. In this study, ViT, VGG and ResNet deep learning networks are used for feature extraction of SVE images to obtain richer information and higher-level feature representations of retinal fundus images.

## *2.5. Classification Model establishment*

### *2.5.1. Models based on images*

Vision Transformer (ViT)[43] is a deep neural network architecture that has been shown to achieve highly competitive performance in a wide range of vision applications such as image classification, object detection, and semantic image segmentation. When pre-trained with enough data, ViT outperforms convolutional neural networks (CNNs), breaks through the limitation of the transformer's lack of an inductive bias, and can achieve better migration results in downstream tasks. In this study, we use the vit-base-patch16-224 pre-trained model for transfer learning to classify 2D retinal fundus images.

Residual Network (ResNet)[44] is a deep neural network architecture that has made significant breakthroughs in the field of computer vision. Its core innovation is the introduction of Residual Block. Jump-connected multilayer networks can form a Residual Block, and ResNet is formed by stacking Residual Blocks. Gradient vanishing leads to deep networks not being able to learn features efficiently, while gradient explosion may lead to unstable model training. With residual blocks, ResNet can solve the gradient vanishing and gradient explosion problems of traditional deep neural networks, and alleviate the network degradation problem caused by deepening network layers.

Visual Geometry Group (VGG)[45] is a deep neural network architecture. The core of VGG lies in the use of a uniform convolutional kernel size i.e., all 3x3 convolutional kernels and 2x2 maximum pooling layers are used. This consistency simplifies the structure of the network and makes it easier to understand and implement. The use of smaller size convolutional kernels (3x3) helps to reduce the number of parameters while providing a larger sense field, allowing the network to capture a wider range of features. VGG makes the entire network more modular by using a uniform structure, while improving the extraction of complex features by increasing the depth of the network.

The above three network architectures were shown in Figure.3.

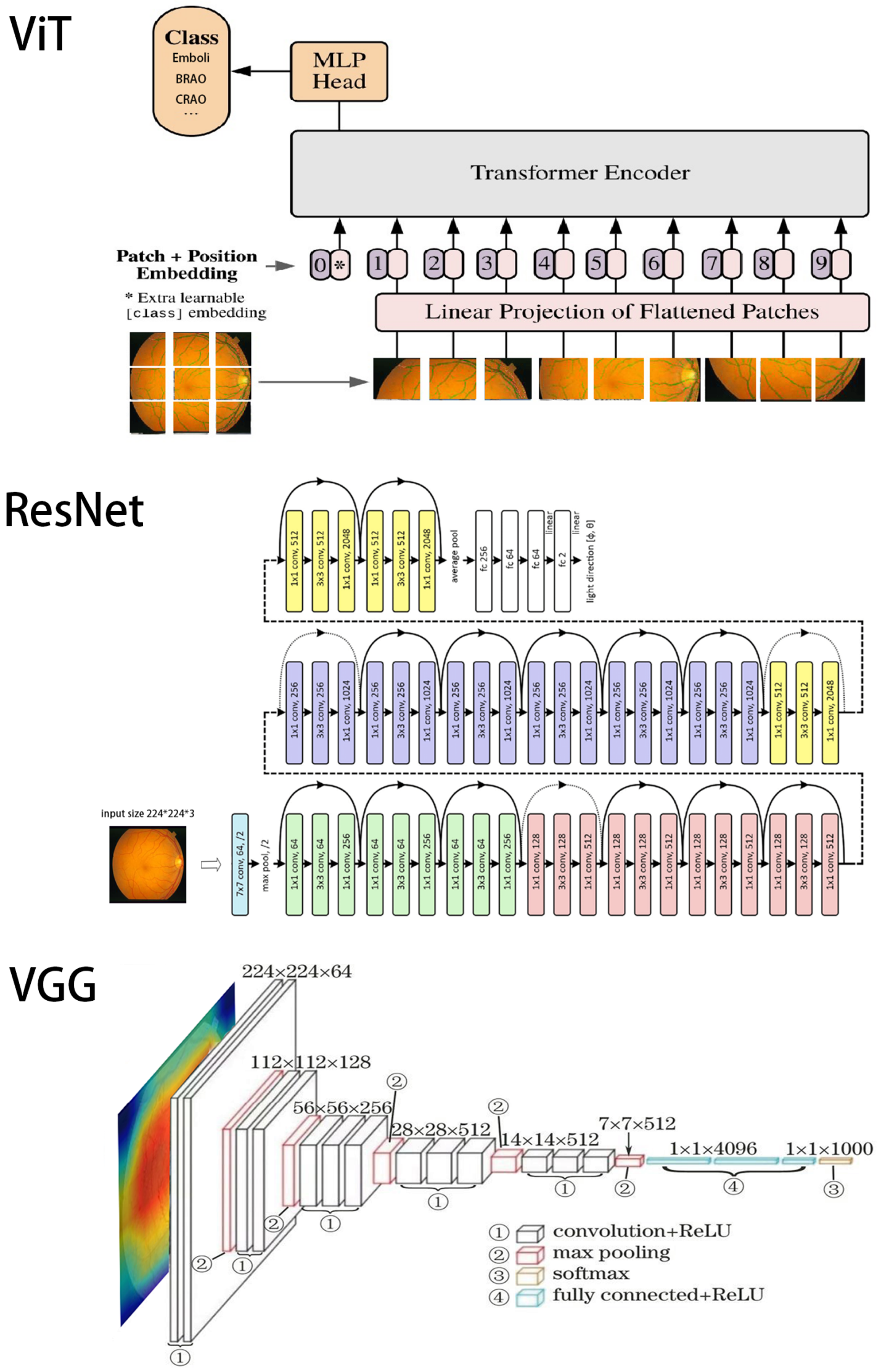

**Figure.3. The rough structure diagrams of ViT, ResNet and VGG.**

### *2.5.2. Models based on features*

When using features as input, this study compared 8 models: SVM, RF, LR, KNN, MLP, 1DResNet, LDA, PLS-LDA in order to get the best model based on this task. SVM(Support Vector Machine)[46] is a commonly used supervised learning algorithm that finds a hyperplane by mapping the input data points into a high-dimensional space, allowing different classes of data points to be separated in this space. RF (Random Forres) [47]is an ensemble learning method, for the classification problem, Random Forest uses the "voting" method, that is, each decision tree votes for a category, and the final prediction is the category that receives the most votes.LR (Logistic regression) [48] is a classical linear classification algorithm, which first fits a decision boundary and then establishes a probabilistic link between this boundary and the classification, thus obtaining the probability of classification labels. KNN(K Nearest Neighbors)[49] is an instance-based classification algorithm, whose core idea is to assume that samples that are similar in feature space have similar classification labels, and therefore new samples are assigned to the closest of K samples belonging to the majority of the categories. 1DResNet (1-Dimension Residual Network) [50]is a simple variant of ResNet, which is obtained by converting the two-dimensional convolutional kernel of ResNet into a one-dimensional convolutional kernel. LDA( Latent Dirichlet Allocation) [51]is a method used for pattern recognition and statistical classification, the goal of which is to find a projection that maps the data from the original high-dimensional space to a low-dimensional space, in order to maximize the differences between categories while minimizing the differences within categories. PLS-LDA(Partial Least Squares Discriminant Analysis ) [52]combines the ideas of PLSR and LDA.it not only takes into account the relationship between categories, but also makes full use of the relationship between the independent variables.

## *2.6. Model assessment*

### *2.6.1. Confusion matrix*

A confusion matrix is a popular model evaluation metric. The horizontal axis in the confusion matrix chart represents the types of predicted results for the samples, while the vertical axis represents the types of labels for the samples. Accuracy, specificity, and sensitivity closer to 1 indicate better diagnostic results of the constructed model. TPR represents sensitivity, FPR represents specificity, TP represents the number of positive samples in the validation set that the model correctly classified, FN represents the number of misclassified positive samples, FP represents the number of misclassified negative samples in the validation set, and TN represents the number of negative samples in the validation set.

In this study, for each class, Precision, Recall, Specificity, F1-Score, Accuracy were calculated. The weighted Average on Precision, Recall, Specificity, F1-score, and Overall Accuracy were also calculated for all classes.

$$Precision = \frac{TP}{TP+FP} \tag{6}$$

$$Recall = \frac{TP}{TP+FN} \tag{7}$$

$$Specificity = \frac{TN}{TN+FP} \tag{8}$$

$$F1-score = \frac{2\times Precision\times Recall}{Precision+Recall} \tag{9}$$

$$Accuracy = \frac{TP+TN}{TP+TN+FP+FN} \tag{10}$$

$$Overall\ Accuracy = \frac{\sum_{i=0}^{classes\ number} TP}{All\ samples} \tag{11}$$

$$weighted\ Avg\ = \sum_{i=0}^{classes\ number} class\{i\}\ metrics \times \frac{class\{i\}\ number}{all\ samples\ number} \tag{12}$$

$$macro\ Avg\ = \frac{\sum_{i=0}^{classes\ number} class\{i\}\ metrics}{classes\ number} \tag{13}$$

#### *2.6.2. ROC curve and AUC*

Receiver Operating Characteristic (ROC) curve and Area Under the Curve (AUC) are important tools for evaluating the performance of classification models. The ROC curve displays the trade-off between True Positive Rate (TPR) and False Positive Rate (FPR) at different classification thresholds, helping us understand the model's performance at various thresholds. AUC is the area under the ROC curve, usually ranging between 0.5 and 1. A higher AUC value indicates better model performance, while a value closer to 0.5 indicates poorer performance. It is a concise yet powerful metric for comparing the performance of different models, with a larger AUC typically indicating that the model can more accurately distinguish between positives and negatives.

## 3. Results and discussion

### *3.1. Data augmentation*

The distribution of samples in Train, Validation, and Test datasets before and after data augmentation is shown in Figure.4. It can be observed that the number of samples in each category is more evenly distributed after data augmentation.

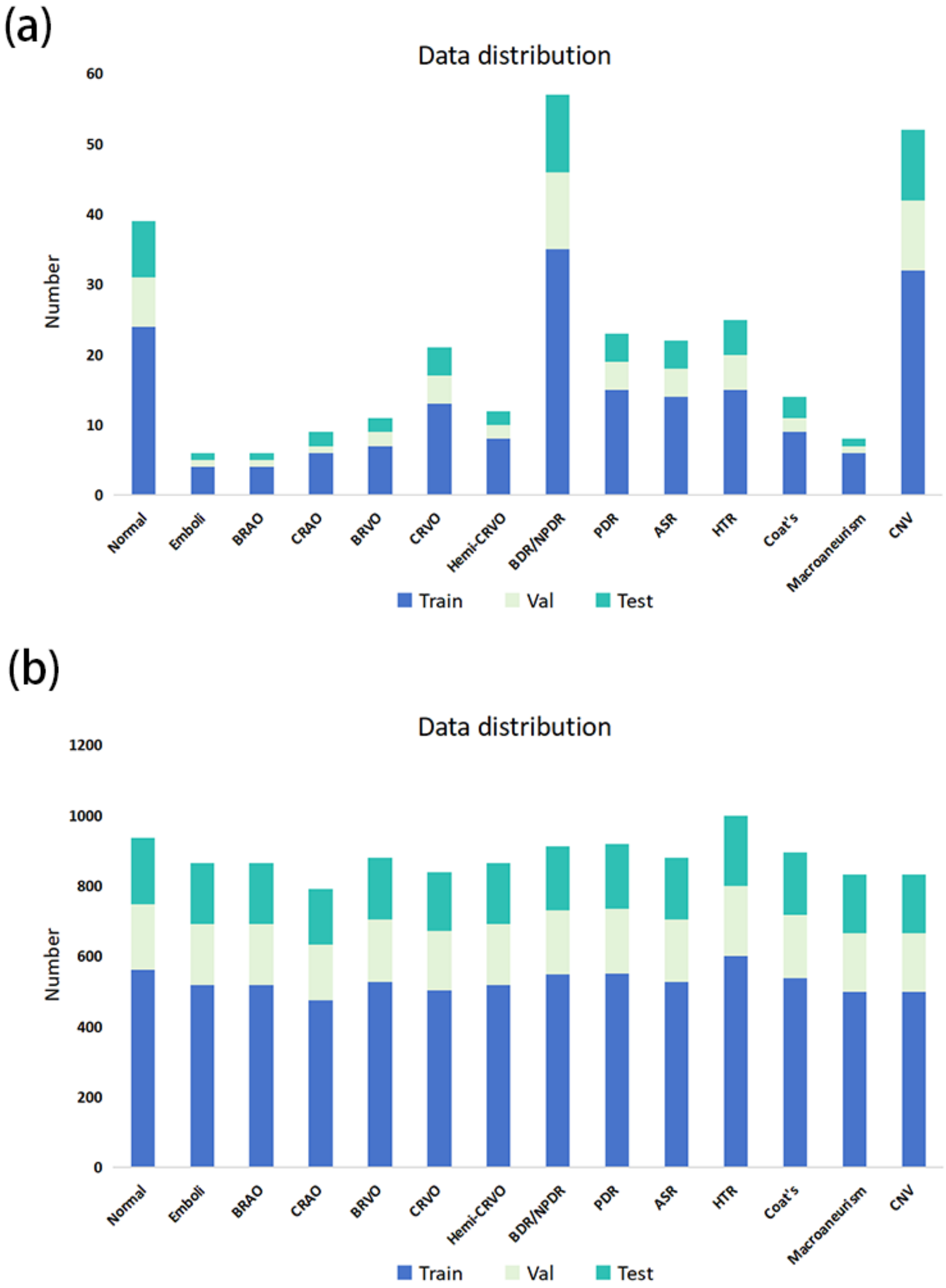

**Figure.4. The samples distribution of training set, validation set and test set.** (a)before augmentation. (b)after augmentation. 6:2:2 ratio was used to randomly divide the training set, validation set and test set. It can be observed that the number of samples in each category is more evenly distributed after data augmentation.

### *3.2. Classification performance*

The classification performances of three DL models on test set for images with different preprocessing methods are shown in Table.3, where all the metrics are weighted Avg except Accuracy which is Overall Accuracy.

| *Preprocessing* | *Model* | *Precision* | *Recall* | *Specificity* | *F1-score* | *Accuracy* | *AUC* |
|---|---|---|---|---|---|---|---|
| *RAW* | *ViT* | *0.9163* | *0.9094* | *0.9548* | *0.9068* | *0.9094* | *0.9512* |
| | *VGG* | *0.9497* | *0.9476* | *1.0000* | *0.9465* | *0.9476* | *0.9718* |
| | *ResNet* | *0.9204* | *0.9139* | *0.9921* | *0.9125* | *0.9139* | *0.9536* |
| *Smooth Grad-CAM++* | *ViT* | *0.8290* | *0.8291* | *0.9803* | *0.8276* | *0.8291* | *0.9079* |
| | *VGG* | *0.9331* | *0.9304* | *1.0000* | *0.9283* | *0.9304* | *0.9625* |
| | *ResNet* | *0.9518* | *0.9504* | *1.0000* | *0.9500* | *0.9504* | *0.9733* |
| *SVE* | *ViT* | *0.9285* | *0.9280* | *0.9763* | *0.9271* | *0.9280* | *0.9612* |
| | *VGG* | *0.9640* | *0.9630* | *1.0000* | *0.9629* | *0.9630* | *0.9801* |
| | *ResNet* | *0.9619* | *0.9614* | *1.0000* | *0.9611* | *0.9614* | *0.9792* |

**Table.3. The classification performances of three DL models on test set for images with different preprocessing methods.** It can be observed that the best results are achieved after using SVE preprocessing.

The classification performances of eight ML models on test set for SVE images with different feature extraction methods are shown in Table.4, where all the metrics are weighted Avg except Accuracy which is Overall Accuracy. From the classification results of all the models for all classes, it can be seen that all the three models, ViT, VGG and ResNet, perform better in classifying the SVE processed images in when using images as input. KNN has the best classification performance when using HOG, LBP and ViT features extracted from SVE images and MLP has the best classification performance when using VGG and ResNet features extracted from SVE images.

| *Preprocessing* | *Feature Extraction* | *Model* | *Precision* | *Recall* | *Specificity* | *F1-score* | *Accuracy* | *AUC* |
| --- | --- | --- | --- | --- | --- | --- | --- | --- |
| *SVE* | *HOG* | *SVM* | *0.5606* | *0.4963* | *0.9778* | *0.4885* | *0.4963* | *0.7285* |
| | | *RF* | *0.0813* | *0.0818* | *1.0000* | *0.0284* | *0.0818* | *0.5033* |
| | | *LR* | *0.0213* | *0.1318* | *0.8289* | *0.0361* | *0.1318* | *0.5323* |
| | | ***KNN*** | ***0.7566*** | ***0.7144*** | ***0.9412*** | ***0.7208*** | ***0.7144*** | ***0.8462*** |
| | | *MLP* | *0.6084* | *0.3865* | *0.7385* | *0.3910* | *0.3865* | *0.6696* |
| | | *PLS-LDA* | *0.2949* | *0.2099* | *0.9359* | *0.1798* | *0.2099* | *0.5739* |
| | | *1DResNet* | *0.4539* | *0.2962* | *1.0000* | *0.2919* | *0.2962* | *0.6198* |
| | | *LDA* | *0.6511* | *0.3047* | *0.9024* | *0.3294* | *0.3047* | *0.6257* |
| | *LBP* | *SVM* | *0.1914* | *0.2032* | *0.9444* | *0.1488* | *0.2032* | *0.5691* |
| | | *RF* | *0.3150* | *0.1085* | *1.0000* | *0.0672* | *0.1085* | *0.5177* |
| | | *LR* | *0.0119* | *0.0711* | *0.0107* | *0.0108* | *0.0711* | *0.5004* |
| | | ***KNN*** | ***0.7936*** | ***0.7899*** | ***0.9796*** | ***0.7894*** | ***0.7899*** | ***0.8868*** |
| | | *MLP* | *0.4536* | *0.4357* | *0.8652* | *0.4153* | *0.4357* | *0.6966* |
| | | *PLS-LDA* | *0.3117* | *0.3039* | *0.9298* | *0.2841* | *0.3039* | *0.6246* |
| | | *1DResNet* | *0.5905* | *0.5773* | *0.9633* | *0.5783* | *0.5773* | *0.7724* |
| | | *LDA* | *0.3302* | *0.3056* | *0.9740* | *0.2823* | *0.3056* | *0.6254* |
| | *ViT (pre_logits layer)* | *SVM* | *0.9392* | *0.9353* | *0.9759* | *0.9337* | *0.9353* | *0.9652* |
| | | *RF* | *0.3523* | *0.1786* | *0.9649* | *0.1671* | *0.1786* | *0.5557* |
| | | *LR* | *0.8655* | *0.8104* | *1.0000* | *0.8163* | *0.8104* | *0.8977* |
| | | ***KNN*** | ***0.9964*** | ***0.9963*** | ***1.0000*** | ***0.9963*** | ***0.9963*** | ***0.9980*** |
| | | *MLP* | *0.9955* | *0.9955* | *1.0000* | *0.9955* | *0.9955* | *0.9976* |
| | | *PLS-LDA* | *0.5288* | *0.5301* | *0.8810* | *0.5236* | *0.5301* | *0.7467* |
| | | *1DResNet* | *0.9843* | *0.9841* | *1.0000* | *0.9841* | *0.9841* | *0.9914* |
| | | *LDA* | *0.9777* | *0.9772* | *1.0000* | *0.9773* | *0.9772* | *0.9877* |
| | *VGG (features layer)* | *SVM* | *0.9960* | *0.9959* | *1.0000* | *0.9959* | *0.9959* | *0.9978* |
| | | *RF* | *0.6863* | *0.3963* | *1.0000* | *0.4176* | *0.3963* | *0.6734* |
| | | *LR* | *0.9756* | *0.9731* | *1.0000* | *0.9733* | *0.9731* | *0.9855* |
| | | *KNN* | *0.9988* | *0.9988* | *1.0000* | *0.9988* | *0.9988* | *0.9993* |
| | | ***MLP*** | ***0.9996*** | ***0.9996*** | ***1.0000*** | ***0.9996*** | ***0.9996*** | ***0.9998*** |

| | | | | | | | |
|---|---|---|---|---|---|---|---|
| | *PLS-LDA* | *0.8436* | *0.8405* | *0.9869* | *0.8392* | *0.8405* | *0.9141* |
| | *1DResNet* | *0.9253* | *0.9219* | *1.0000* | *0.9228* | *0.9219* | *0.9579* |
| | *LDA* | *0.9631* | *0.9601* | *1.0000* | *0.9610* | *0.9601* | *0.9786* |
| *ResNet (avgpool layer)* | *SVM* | *0.9952* | *0.9951* | *1.0000* | *0.9951* | *0.9951* | *0.9974* |
| | *RF* | *0.8433* | *0.5980* | *1.0000* | *0.6057* | *0.5980* | *0.7825* |
| | *LR* | *0.9659* | *0.9630* | *1.0000* | *0.9632* | *0.9630* | *0.9800* |
| | *KNN* | *0.9980* | *0.9980* | *1.0000* | *0.9980* | *0.9980* | *0.9989* |
| | ***MLP*** | ***0.9984*** | ***0.9984*** | ***1.0000*** | ***0.9984*** | ***0.9984*** | ***0.9991*** |
| | *PLS-LDA* | *0.8905* | *0.8897* | *1.0000* | *0.8889* | *0.8897* | *0.9406* |
| | *1DResNet* | *0.9964* | *0.9963* | *1.0000* | *0.9963* | *0.9963* | *0.9980* |
| | *LDA* | *0.9968* | *0.9967* | *1.0000* | *0.9967* | *0.9967* | *0.9982* |

**Table.4. The classification performances of eight ML models on test set for SVE images with different feature extraction methods.** It can be observed that among all the models, SVE-VGG-MLP is optimal across all categorical metrics.

SVE-VGG-MLP has the best classification performance with a weighted Average on Precision, Recall, Specificity, F1-score, Overall Accuracy and AUC of 0.9984, 0.9984, 1.0000, 0.9984, 0.9984 and 0.9991 respectively. The confusion matrix and ROC curve for the classification of this model are shown in Figure.5.

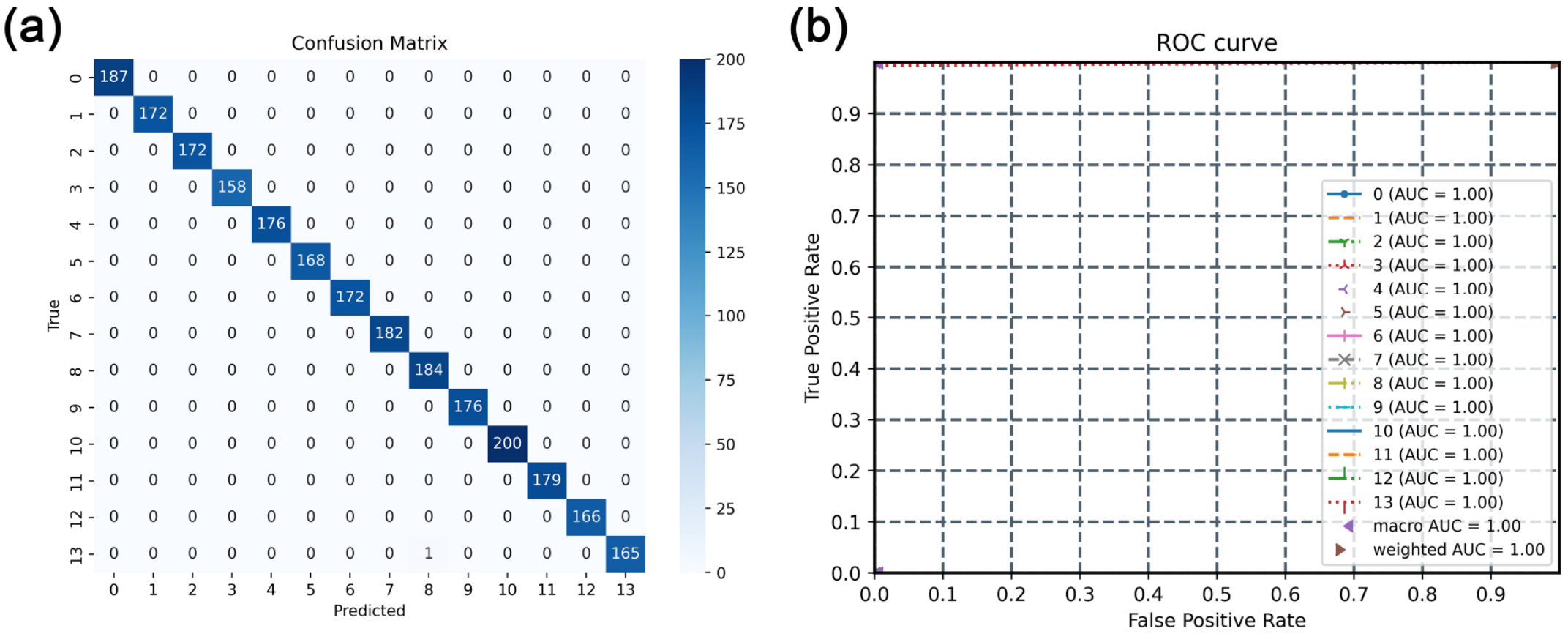

**Figure.5. The classification result of SVE-VGG-MLP model.** (a)The confusion matrix. (b)The ROC curve.

This study also validated the classification performance of different models for different diseases. The AUC of all models for different diseases, macro AUC and weighted AUC are shown in Figure.6.

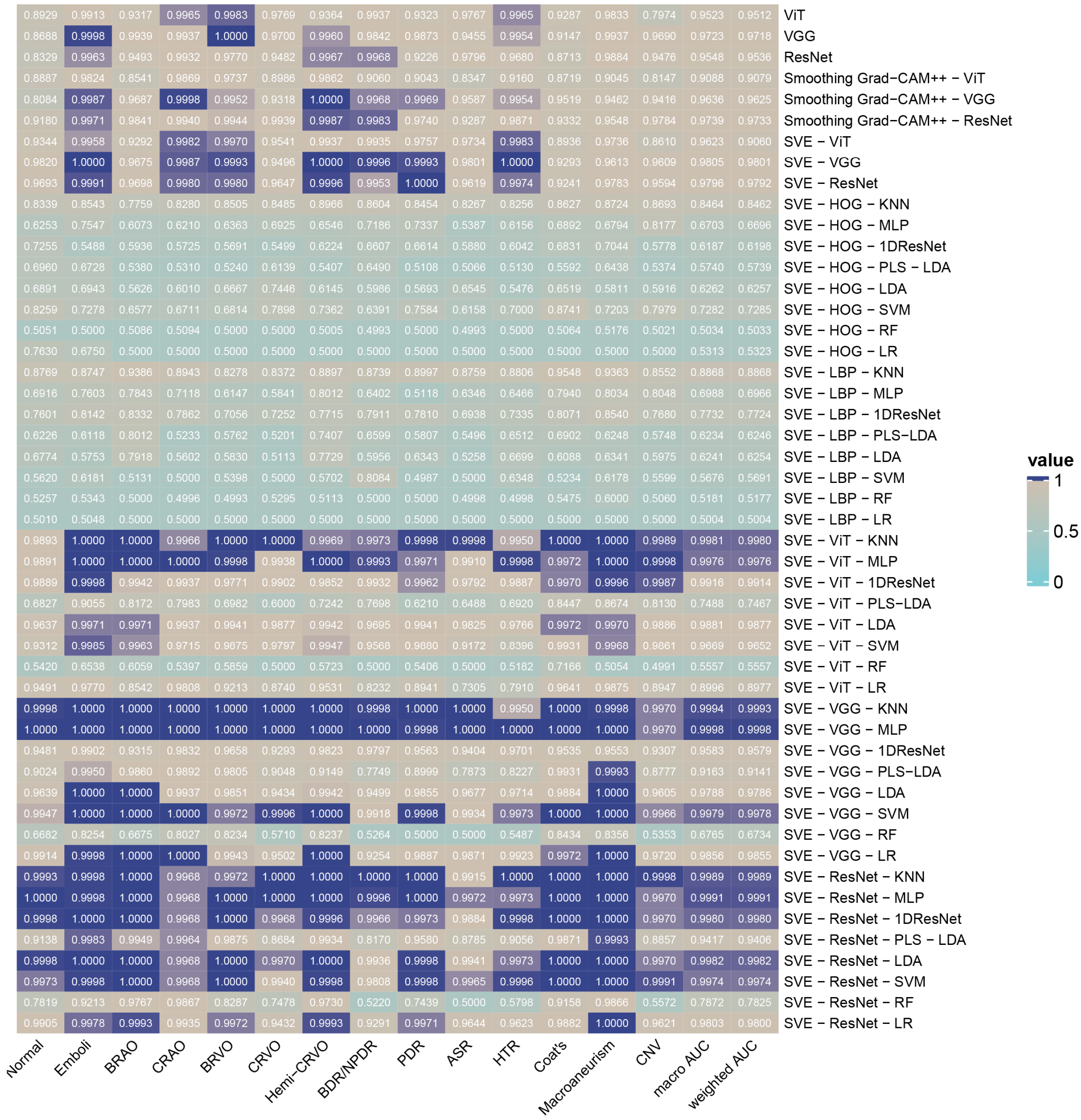

**Figure.6. The AUC heatmap of different models for different diseases.** The classification performance of different models for different diseases can be observed, among which the SVE-VGG-MLP model has the best classification performance, with the AUC equal to 1 for almost all categories, and only CNV has an AUC of 0.997.

# 4. Conclusion

Analyzing vascular structures in retinal fundus images is significant for the diagnosis of a wide range of diseases. In this study, we used diverse Data Augmentation strategies and proposed a novel SVE preprocessing method to acquire SVE images and compared the classification result with Origin images and Smooth

Grad-CAM ++ images to evaluate the classification ability of different preprocessing methods for diseases related to retinal fundus images. For the SVE images, we extracted their deep learning features and traditional features, and established several 1D models to evaluate the classification ability of different feature extraction methods and different meta learners for retinal fundus images related diseases. For 2D images processed by different method, the SVE preprocessing method proposed in this study always gives the best prediction on different models. The SVE-VGG-MLP model established with deep learning features extracted from SVE images has best classification ability. The whole study shows that combining SVE and deep learning features can more accurately classify retinal fundus images related diseases, which provides a new method for the detection and classification of retina-related diseases.

## Acknowledgements

The authors would like to thank all the reviewers who participated in the review. We also acknowledge CHASE DB1, DRIVE, HRF and STARE databases for providing their platforms and contributors for uploading their meaningful datasets.

## Ethical Statement

Not applicable.

## Data Availability

The datasets used in this study are all public and freely available. Codes will be made available on request.

## Funding

This research did not receive any specific grant from funding agencies in the public, commercial, or not-for-profit sectors.